# QUANTUM ASTRONOMY

By Alexander M. Ilyanok

## PART I

### Energy of stars and the Hollow Sun
### (nonthermonuclear approach)

*The article describes new model of the Sun having a hollow core as introduced in [1]. The model helps to explain a number of experimental facts of kinetics, energetics, and Sun spectroscopy based on classic physics. The origin of the Sun's energy is not the thermonuclear process taking place in its core, but the coherent, anisotropic, gravitational compression of the atomic hydrogen in the solar shell at the temperature of 6289 K.*

«*Иллюзии гибнут – факты остаются*».
   *Д.И. Писарев*

В статье развивается новая модель Солнца с полостью внутри, представленная в работе [1]. На основе классической физики с помощью этой модели объяснен ряд экспериментальных фактов по кинетике, энергетике и спектроскопии Солнца. Показано, что источником энергии Солнца является не термоядерный синтез в его ядре, а когерентное анизотропное гравитационное сжатие атомарного водорода в солнечной оболочке с температурой 6293К.

На каждом этапе формирования новых парадигм в физике возникали адекватные модели, описывающие энергию звезд, в том числе и энергию Солнца. Так, например, в XIX в. Гельмгольц считал основным источником энергии Солнца энергию гравитационного сжатия [2], которое приводит ко времени полного выгорания Солнца:

$$t_0 = \frac{W_0}{L_0} = \frac{GM_0^2}{L_0 R_0} = 33 \text{ млн. лет,} \tag{1}$$

где $W_0$ и $L_0$ – гравитационная энергия и полная мощность излучения Солнца соответственно; $M_0$ и $R_0$ – масса и радиус Солнца соответственно; $G$ – гравитационная постоянная.

Однако последующие палеонтологические исследования показали, что в течение последних 3-х миллиардов лет жизнь на Земле не прерывалась и, следовательно, светимость Солнца существенно не менялась. Этот научный факт привел к первому кризису в физике энергетики звезд.

Преодолеть этот кризис уже в XX в. пытались многие ученые, в том числе Эддингтон (1926). Он нашел, что для основных звезд достаточно в грубом приближении светимость связана с массой звезды в пропорции

$$L \sim M^\gamma, \tag{2}$$

где $\gamma = 3 \div 3.5$.

При такой зависимости теплопроизводительность звезд определяется только теплоотдачей. Таким образом, механизм выделения энергии звездами соответствует типу выделения энергии при остывании или освобождении гравитационной энергии при сжатии.

Пытаясь устранить недостатки теории Гельмгольца, Эддингтон ввел новый источник энергии в виде равномерной аннигиляции вещества (в соответствии с уравнением Эйнштейна) во всем объеме звезды. Однако дальнейшее развитие теории и эксперимента в ядерной физике показало, что при существующих физических условиях на Солнце процесс аннигиляции не возможен. Это привело ко второму кризису в физике энергетики звезд.

Осмысливая создавшееся положение, большинство физиков склонилось к идее термоядерного механизма выделения энергии в звездах. Вскоре Бете и др. (1938-1939) указали на наиболее важные термоядерные реакции протон-протонного цикла, углеродно-азотного цикла и углеродного цикла, которые могут осуществляться при давлениях порядка $10^{16}$ Па и температурах порядка $10^8$ К. Получить такие давления и температуры возможно только в центре звезд. Но эта модель вошла в противоречие с экспериментальной зависимостью (2) светимости звезд от их массы. Поэтому было создано множество теорий, пытающихся устранить это противоречие.

В начале 70-х годов всеобщая уверенность в термоядерном механизме генерирования солнечной энергии была поколеблена тем фактом, что непосредственно измеренный поток солнечных нейтрино, выделяемых при термоядерном синтезе, оказался в 3-4 раза меньше теоретически ожидаемого. Это привело к третьему кризису в физике энергетики звезд. Он получил название проблемы солнечных нейтрино. Неоднократные экспериментальные попытки на протяжении последующих 30 лет найти теоретически предсказанное количество солнечных нейтрино и выделить их на общем космическом фоне оказываются безуспешными до настоящего времени.

Таким образом, кризис в физике энергетики звезд перешел в хроническую форму.

Можно попытаться преодолеть этот кризис, полностью отказавшись от стандартных физических моделей Солнца и звезд, перейдя к модели, учитывающей вновь полученные данные по физике Солнца и звезд [1].

Новые наблюдения и факты по физике Солнца ставят под сомнение общепризнанные теории его строения. Так, в 1975 г. в СССР [3], и США [4], независимыми группами ученых были открыты пульсации солнечной фотосферы с периодом 160.01 мин и амплитудой 6 км. С точки зрения теории адиабатических пульсаций газовых шаров этот результат говорит о том, что плотность Солнца не возрастает по мере приближения к центру, как это принято считать, а убывает. Это сопровождается весьма незначительным возрастанием температуры к центру [2] до $T_c \leq 6.5 \cdot 10^6$ К. При таких температурах протон-протонная реакция даст энергии в $10^4$ раз меньше, чем фактически наблюдается у Солнца.

В 1960 г. Лейтон обнаружил, что вся поверхность спокойного Солнца (фотосфера и начало хромосферы) покрыто областями, колеблющимися по вертикали с периодом около 5 мин. [5]. Эти колебания представляют синусоидальные волновые цуги длительностью в 4-5 циклов и со средней продолжительностью 23 мин. В последующих работах стало предполагаться, что множество областей поверхности Солнца размером с ячейку супергрануляции колеблются когерентно. Горизонтальные масштабы амплитудной когерентности лежат в пределах 5-10 тыс. км, в то время как фазовая когерентность сохраняется до размеров 30 тыс. км.

Связи между 160 и 5 минутными колебаниями выявлены не были. Это создало дополнительные проблемы интерпретации 160 минутных сейсмических колебаний.

Эти проблемы стимулируют поиски нестандартных моделей строения Солнца, которые способны были бы объяснить имеющиеся экспериментальные данные.

Интересный опыт был поставлен космонавтами Б. В. Волыновым и В.И. Жолобовым по изучению поведения газовых пузырей в жидкости в условиях невесомости [6]. Сферическая колба диаметром 3 см заполнялась водой, а затем интенсивно встряхивалась, В ней образовывалось множество воздушных пузырьков. Примерно через 100 ч. в колбе оказалась одна сферическая полость, расположенная в воде почти в центре колбы. Таким образом, в ансамбле газонаполненных пузырьков, хаотически распределенных в воде в условиях невесомости, происходит процесс их объединения, который завершается формированием одного крупного пузыря. Для звезды роль сферической колбы могут выполнять гравитационные силы. Поэтому модель с разреженной центральной частью (полостью) внутри звезды может оказаться приемлемой для Солнца.

Предположение о наличии полости в Солнце сочетается с некоторыми гипотезами, имеющимися в литературе. Например, в работе [7] есть интересное предположение, вытекающее из принципов симметрии. Из них следует, что относительно планетной системы Солнца внутри самого Солнца также должны быть некие «зеркальные» планеты.

Распределение массы внутри Солнца можно было бы установить, измеряя его момент инерции. Прямые эксперименты по измерению момента инерции Солнца и распределения его массы можно провести аналогично определению распределения массы внутри планет, имеющих спутники. Обычно такие вычисления следуют из движения перигелия и узлов спутников планет относительно орбиты самой планеты. Так как период обращения Солнца вокруг центра галактики составляет 220 млн. лет, то для проведения прямых экспериментов по вращению перигелиев и узлов планет относительно орбиты Солнца требуется, по крайней мере, миллионы лет. Поэтому нельзя этим прямым методом определить распределение массы внутри Солнца. Единственный довод, который приводится в пользу большей концентрации материи к центру звезд – это медленное движение линии абсид тесных звездных пар [2]. Однако более подробная сводка движения для линий абсид ряда двойных звезд показывает полное отсутствие корреляции между скоростью этого движения и отношением радиуса звезды к полуоси орбиты. Эта корреляция должна быть обязательна, если движение абсид связано с деформацией фигур гравитирующих тел.

Прямым методом остается только расчет распределения массы внутри Солнца по его сжимаемости вследствие вращения.

Решение для Солнца задачи вращения в общем виде при произвольной форме уравнения состояния вещества планеты невозможно провести аналитически. При достаточно малых деформациях эту задачу можно решить приближенно путем разложения деформаций, вызванных вращением, в виде сходящихся функциональных рядов. Применение теории малых деформаций позволяет найти рациональное обоснование форм фигур равновесия и установить их связь с внутренним строением небесных тел.

Одним из методов приближенного решения проблемы относительного равновесия небесных тел является метод Ляпунова [8]. Проблема малых деформаций в его постановке ограничена отысканием фигур равновесия, близких к какой-либо заранее заданной форме геометрической поверхности.

Наибольший интерес заслуживает решение Ляпуновым задачи Клеро - определение фигуры равновесия медленно вращающейся неоднородной планеты.

Рассмотрим Солнце как идеальную жидкость. При очень малой угловой скорости деформация здесь будет незначительной, и фигура равновесия окажется близкой к сфере, которую можно выразить уравнением сфероида Клеро:

$$r = a[1 - \sigma \sin^2\varphi],$$

где $\varphi$ - гелиоцентрическая широта точки поверхности; $\sigma$ - сжатие фигуры равновесия, определяемое как

$$\sigma = (a-b)/a,$$

где $a$ и $b$ - большая и малая полуоси фигуры.

Представим Солнце в виде медленно вращающегося шара, состоящего из сжимаемой жидкости. При этом смоделируем сжимаемость жидкости в виде двух компонент: оболочка Солнца состоит из жидкости с плотностью $\rho_1$, а ядро из жидкости с плотностью $\rho_2$. Для медленно вращающегося тела, находящегося в гидростатическом равновесии и симметричного относительно оси вращения и экваториальной плоскости гравитационный потенциал $V$ можно разложить в ряд [9]. В этом случае сжатие фигуры равновесия Солнца определяется только четными членами ряда, начиная с $n \geq 4$. Оценивая только первый член ряда, воспользуемся известными выводами [8]. Из этой работы следует, что сжатие $\sigma$ фигуры равновесия по задаче Клеро методом Ляпунова определяется равенством:

$$\frac{\sigma}{4\pi M} - \frac{\sigma}{5R^2}\int_{r_1}^{R}\rho(r)r^4 dr = \frac{\omega^2 R^3}{8\pi G}, \qquad (3)$$

где $R$ - радиус невозмущенной звезды; $r_1$ – радиус полости.

Производя интегрирование (3) при равномерном распределении плотности $\rho(r) = $ const получаем

$$\sigma = \frac{\omega^2 R^3}{2MG}\left[1 - \frac{3}{5}\frac{1-(r_1/R)^5}{1-(r_1/R)^3}\right]^{-1}. \qquad (4)$$

Если $r_1 = 0$, то есть масса равномерно распределена по объему Солнца, то:

$$\sigma = (5/4)(\omega^2 R^3)/(MG) = 2.6\cdot10^{-5}, \qquad (5)$$

что соответствует классическому решению Ньютона.

Если масса сосредоточена в центре Солнца, то из (3) следует:

$$\sigma = (\omega^2 R^3)/(2MG) = 1.04\cdot10^{-5}. \qquad (6)$$

Обе эти модели не соответствуют экспериментально измеренному значению сжимаемости Солнца, равному $5.21 \cdot 10^{-5}$ [10]

Подставив в (4) экспериментальное значение сжатия Солнца находим, $r_1/R = 0.763$.

Этот результат показывает, что уже в первом приближении единственным решением для сжатия Солнца является перераспределение его основной массы на его оболочку. Учитывая большее количество четных членов ряда по аналогии с работой [9], получаем асимптотическую предельную толщину оболочки солнца из $r_1/R = 0.962$, т.е. толщина оболочки Солнца составляет $\Delta R_0 = R_0/26.6 = 2.61845 \cdot 10^7$ м. К аналогичным результатам можно придти, используя квантово-механическую модель полого Солнца. В дальнейшем будем предполагать, что полость Солнца заполнена газообразным водородом в виде низкотемпературной плазмы, имеющей давление, соизмеримое с давлением на видимой поверхности Солнца, которое составляет порядка 0.1 атм. [10].

Так как предполагается, что Солнце представляет собой сферу с полостью внутри, то гелиосейсмическую волну с периодом 160.01 мин. можно рассматривать как колебание самой оболочки. Тогда, если предположить, что распространение гравитационных волн на Солнце происходит по внутренней стороне оболочки Солнца и скорость их движения не превышает первой космической скорости на поверхности Солнца, равной $v_1 = 437$ км/с [10], то время распространения такой волны вдоль оболочки составит

$$t_1 = 2\pi R/v_1 == 160.43 \text{ мин.} \qquad (7)$$

при $R = R_o(1 - 1/26.6)$ получается расхождение с экспериментальным значением всего 0.26%.

Если предположить, что поперек оболочки Солнца также распространяются волны со скоростями $v_n = v_1/(2n+1)$, то для второй моды $n=2$ получаем время прохождения волны между внутренней и внешней поверхностью оболочки

$$t_2 = 5\Delta R_0/v_1 = 5.00 \text{ мин.} \qquad (8)$$

Прямым фактом, подтверждающим наличие тонкой оболочки толщиной $\Delta R_0 = 2.61845 \cdot 10^4$ км, является существование долгоживущих (порядка 20 часов) ячеек супергрануляции с характерными размерами $2 \cdot 10^4 \div 4 \cdot 10^4$ км. Эти ячейки напоминают синергетические ячейки Бенара (Benard), возникающие в любом плоском горизонтальном слое вязкой жидкости при подогреве снизу. Упорядоченное поведение таких ячеек происходит за счет пространственных корреляций при их взаимодействии. При этом размер ячеек соизмерим с толщиной слоя вязкой жидкости и определяется из следующих характерных габаритных размеров сосуда: отношение глубины $D$ к ширине/длине $L$ сосуда должно быть $10 < L/D < 30$. Для Солнца отношение $L/D = R_0/\Delta R_0 = 26.6$. Причем время существования таких ячеек порядка суток, что совпадает со временем существования ячеек супергрануляции [11] Поразительно, но факт, что когерентные анизотропные процессы, идущие в сосуде с ртутью или маслом аналогичны процессам, идущим в солнечной оболочке, хотя масштабы этих процессов несоизмеримы. В работе [11] приведена иллюстрация возникновения когерентных анизотропных процессов, возникающих в жидкой металлической сферической оболочке при ее резком охлаждении при сбросе внутреннего давления. Полученные регулярные структуры на поверхности металлической оболочки очень напоминают ячейки супергрануляции на Солнце, что еще раз подтверждает правомерность выбранной модели Солнца.

Кроме того, на Солнце не найдено общее магнитное поле, аналогичное земному, что говорит об отсутствии у него ядра. Однако существует большое количество локальных магнитных полей, связанных с ячейками супергрануляции и солнечными пятнами. Это свидетельствует о наличии вихревых токов в ячейках супергрануляции [10].

Можно показать, что наличие полости в Солнце следует и из законов сохранения кинетической энергии его вращения и кинетической энергии поступательного движения планет. Парадокс несоблюдения законов полного момента количества движения и полной кинетической энергии солнечной системы обсуждается на протяжении уже не одного столетия.

Со времен Ньютона вопрос о механизме начального толчка движения планет оставался открытым. Только в 1960 году Эдьед (Egyed) на основе старой гипотезы Дирака (Dirac) о изменении гравитационной постоянной со временем нашел решение. Он установил, что первоначально Солнце являлось звездой-гигантом, которая при сжатии периодически сбрасывала свою массу в виде планет, начиная с Плутона, передавая им начальный импульс движения [12].

Аналогично теории Эдьеда предположим, что Солнце в начальный момент было звездой-гигантом, при этом предположим существование внутри него полости. В этом случае на внутренней стенке Солнца будет отсутствовать гравитационный потенциал вследствие зависимости потенциала $1/R$. Следовательно, любое возмущение на внутренней экваториальной поверхности Солнца может породить солитон – каплю материи, которая будет являться планетой. При сжатии оболочки Солнца эта планета сохраняет момент импульса движения и остается на заданной орбите. В этом случае гравитационный потенциал по отношению к планете претерпевает скачок в $4\pi$. Это следует при переходе от уравнения Пуассона внутри гравитирующей оболочки к уравнению Лапласа вне оболочки или аналогично переходу через двойной электрический слой или двойной гравитационный.

Кинетическая энергия Солнца, представленного в виде вращающейся сферы с массой, сосредоточенной в основном в оболочке будет

$$W_k = (M_0 v_3^2)/3 = 2.64 \cdot 10^{36} \text{ Дж}, \qquad (9)$$

где $v_3$ – экваториальная скорость на поверхности Солнца.

Кинетическая энергия движения всех планет солнечной системы составит [10]

$$W_p = \frac{1}{2} \sum_{n=1}^{9} M_n v_n^2 = 1.99 \cdot 10^{35} \text{ Дж}, \qquad (10)$$

где $M_n$ – массы планет, $v_n$ – орбитальные скорости планет.

Приравнивая (9) и (10) получим для закона сохранения кинетической энергии в солнечной системе с учетом скачка гравитационного потенциала в $4\pi$ при переходе через солнечную оболочку следующее выражение:

$$\frac{1}{3} M_0 v_3^2 = 2\pi \sum_{n=1}^{9} M_n v_n^2. \qquad (11)$$

Погрешность при расчете по этой формуле составляет 5.4%, что еще раз свидетельствует о высокой достоверности модели полого Солнца. Закон сохранения энергии солнечной системы (11) показывает, что наша солнечная система не могла сталкиваться с другими звездами после момента ее рождения и не имеет других достаточно крупных планет кроме известных восьми.

Отметим еще два очень важных аргумента, подтверждающих гипотезу о полости в Солнце. Первый - это увеличение яркости края Солнца при наблюдении его в сантиметровом радиодиапазоне. Второй – увеличение красного смещения фотонов к краю Солнца [13]. Это говорит о том, что масса Солнца в основном сосредоточена на его поверхности. Это вызывает неравномерное гравитационное красное смещение оптических фотонов и увеличение концентрации источников СВЧ излучения.

Кроме того, предложенная модель позволяет объяснить вспышки на Солнце в виде протуберанцев. Так как внутри сферически симметричной полости отсутствует гравитационное поле, то там, возможно, образуются небольшие планеты. Они могут двигаться внутри полости, соударяясь со стенками оболочки. При каждом ударе кинетическая энергия переходит в энергию возбуждения оболочки Солнца, которая проявляется как солнечная активность и локальный термоядерный синтез тяжелых элементов.

Но возникает новая проблема – что является источником энергии Солнца?

Предположим, что полое Солнце под действием гравитационных сил находится в состоянии упругого сжатия. Тогда такую систему можно описать дифференциальным уравнением второго порядка по аналогии с упругими полыми сферами, рассматриваемыми в классической механике и находящимися под внешним равномерным давлением. Такие модели хорошо известны, и способы их решения широко представлены в работе [14]. Если представить, что роль внешнего равномерного давления выполняет гравитационное взаимодействие между частицами оболочки Солнца, то дифференциальное уравнение имеет решения в виде: равномерного движения тела; его вращения вокруг оси и волн в оболочке. Все эти движения присущи Солнцу. В последнем случае потенциальная энергия упругого гравитационного сжатия переходит в кинетическую энергию незатухающего движения волн по оболочке.

Представим, что незатухающая волна движется вдоль внутренней оболочки Солнца с первой космической скоростью на поверхности Солнца $v_1 = 437$ км/с. Эта волна вызывает движение электронов и протонов относительно друг друга. Следовательно, кинетическая энергия электрона, двигающегося относительно протона, будет эквивалентна некой температуре:

$$T_e = (m_e v_1^2)/2k = 6293 К, \qquad (12)$$

где $m_e$ – масса свободного электрона, $k$ – постоянная Больцмана.

Вследствие того, что температура по диску Солнца неоднородна и уменьшается к краям из-за увеличения взаимодействия с короной, воспользуемся экспериментальными данными по температуре Солнца, отнесенными к его центру. Экспериментальное значение температуры в центре диска Солнца по спектру со сглаженными неоднородностями составляет 6270К [10]. Таким образом, погрешность между экспериментальным и теоретическим значением составляет 0.31%. Это расхождение, по-видимому, связано с конечной теплопроводностью оболочки Солнца, так как расчетная температура относится к внутренней стороне его оболочки.

Расчет средней плотности солнечной оболочки при данной модели дает 12.97 г/см$^3$, что соответствует плотности ртути при температуре 533К, но в 9.21 раз превышает среднюю по объему плотность Солнца, равную 1.409 г/см$^3$ и в 417 раз превышает плотность жидкого водорода в двойной критической точке 0.0311г/см$^3$. Такая плотность в 12 раз ниже предполагаемой плотности в центре Солнца, равной 160 г/см$^3$[10]. При этой плотности вероятность термоядерного синтеза в оболочке Солнца близка к нулю.

Прямым экспериментальным подтверждением наличия солнечной оболочки такой плотности является совпадение ее плотности с плотностью внешнего ядра Земли. По распространению сейсмических волн установлено, что в Земле существует внешнее и внутреннее ядро. Внешнее ядро начинается на расстоянии 1217.1 км от центра Земли, и его плотность по справочным данным составляет 13.012 г/см$^3$ [10,15]. Погрешность в разнице плотностей оболочки Солнца и ядра Земли составляет 0.3% . Очень важно, что на этих глубинах расчетная температура составляет 6200К ÷ 6300К, что совпадает с температурой оболочки Солнца (12). Кроме того, на протяжении всего внешнего ядра Земли 1217.1–3485.7 км полностью отсутствуют акустические поперечные волны, а существуют только продольные, что характерно только для жидких и газообразных сред. Сейсмоакустические исследования внутреннего ядра дают весьма противоречивые результаты по определению его плотности из-за отражения волн на поверхности раздела внутреннего и внешнего ядер. Здесь также отсутствуют поперечные акустические волны. По-видимому, это говорит о том, что внутренне ядро Земли представляет собой полость, заполненную газообразным водородом при очень высоком давлении (расчетное давление в центре Земли равно 3.63·10$^{11}$ Па [10]), окруженную оболочкой (внешнее ядро) из водородной плазмы в состоянии, аналогичном солнечной оболочке, покрытой мантией.

Таким образом, приходим к весьма интересному выводу: Земля является потухающей водородной звездой. При выгорании водорода синтезируются тяжелые ядра посредством достаточно холодного ядерного синтеза – трансмутации при температуре 6293К - с последующим охлаждением в мантии и кристаллизацией в коре.

Необычное состояние солнечного вещества, которое в общем случае представляет собой низкотемпературную водородную плазму, включающую в себя H$_2$, H, H$^+$, H$^-$, e$^-$ с преобладанием атомарного водорода, требует изменения в представлении *G* как не постоянной, а зависящей от фазового состояния (температуры) вещества. Здесь нет ничего удивительного, так как все эксперименты со времен Кавендиша по измерению *G* проводились с конденсированными веществами в нормальных условиях. Эксперименты по измерению *G* для газов, нейтронов, протонов и электронов проводились как для элементарных частиц [16]. Однако для конденсированного анизотропного вещества эксперименты не проводились. То есть, прямых экспериментов по измерению *G* для вещества Солнца не проводилось. Это предстоит сделать в будущем. Поэтому, в настоящее время нельзя разделить произведение *GM* при исследовании Солнца, а также и Земли при учете фазового состояния его ядра. Следовательно, правильно «взвесить» Землю и Солнце еще предстоит.

Развивая идею Дирака, будем считать, что гравитационная постоянная зависит не столько от времени, сколько от фазового состояния вещества. Предположим, что в начальном состоянии Солнце представляло собой сферу, состоящую, аналогично Юпитеру, из жидкого водорода с критическими

параметрами – плотностью 0.031 г/см$^3$ и критической температурой 32.98К. При его разогреве G вдоль оболочки за счет высокотемпературных когерентных взаимодействий атомарного водорода изменилось в 417 раз, а в поперечном направлении осталось прежним.

Тогда уравнение Гельмгольца (1) будет справедливо, если принять для Солнца $G_0 = 417\ G$. Следовательно, время высвечивания Солнца увеличивается в 417 раз, что составит время его существования 13.76 млрд. лет, что совпадает с оценками времени существования Метагалактики [10].

Однако, эти данные противоречат времени существования Солнца, оцененному по времени возникновения Земли, – примерно 4.5 млрд. лет. Время существования Земли достаточно достоверно определено по соотношению изотопов в метеоритах и земной коре. Но нельзя делать простую интерполяцию полученных на Земле данных как на Солнце, так и на другие звезды без учета разницы их фазового состояния и температуры.

Таким образом, используя модель Солнца с полостью внутри удается объяснить:

1. Наблюдаемое значение сжимаемости Солнца (4);
2. Закон сохранения энергии вращения в солнечной системе (11);
3. Генетическую связь с Солнцем, возникающую при рождении планет;
4. Первоначальный толчок в движении планет солнечной системы;
5. Гелиосейсмические волны с периодом 160 мин. и 5 мин., наблюдаемые на поверхности Солнца (7), (8);
6. Механизм образования ячеек супергрануляции;
7. Отсутствие общего магнитного поля;
8. Вспышки на Солнце;
9. Дефицит солнечных нейтрино как результат отсутствия термоядерного синтеза;
10. Температуру Солнца (12);
11. Гравитационное красное смещение на краях Солнца;
12. Увеличение яркости СВЧ излучения на краях Солнца;
13. Плотность материи Солнца, равную плотности внешнего ядра Земли;
14. Температуру внешнего ядра Земли, равную температуре Солнца;
15. Кубическую зависимость светимости Солнца от его массы (2).

Полученные результаты говорят о высокой достоверности предложенной модели существования полости внутри Солнца.

Таким образом, третий кризис в физике энергетики Солнца может быть преодолен с помощью предлагаемой модели полого Солнца. Из нее следует важный практический вывод: глобального термоядерного синтеза в центре Солнца нет. Источником энергии является особое состояние солнечного вещества в виде когерентного гравитационного анизотропного состояния плазмы, т.е. высокотемпературная гравитационная потенциальная энергия сжатия оболочки переходит в кинетическую энергию вращения и энергию свечения Солнца. Следовательно, по своему характеру энергия на Солнце является чистой и не содержит нейтронных компонент.

Учитывая вышеизложенное, необходимо пересмотреть программы получения энергии на Земле на основе термоядерного синтеза как источника нейтронов, а следовательно, фактора загрязнения окружающей среды и перейти к поиску путей освоения энергии путем трансмутации элементов.